\documentclass[preprint,aps]{revtex4}
\usepackage{graphicx}
\usepackage{dcolumn}
\usepackage{bm}
\begin{document}

\title{Transverse decoherence and coherent spectra
in long bunches with space charge}
\author{Vladimir Kornilov} 
\affiliation{GSI Helmholtzzentrum f\"ur Schwerionenforschung, Planckstr.\,1,
64291 Darmstadt, Germany} 
\author{Oliver Boine-Frankenheim}  
\affiliation{GSI Helmholtzzentrum f\"ur Schwerionenforschung, Planckstr.\,1,
64291 Darmstadt, Germany} 
\affiliation{Technische Universit\"at Darmstadt, Schlossgartenstr.\,8,
64298 Darmstadt, Germany} 

\date{May 7, 2012}

\begin{abstract}
The transverse bunch spectrum and the transverse
decoherence/recoherence following an initial bunch
offset are important phenomena in synchrotrons and storage rings,
and are widely used for beam and lattice measurements.
Incoherent shifts of the particles betatron frequency
and of the synchrotron frequency modify
the transverse spectrum and the bunch decoherence.
In this study we analyze the effects of transverse space charge
and of the rf nonlinearity on the decoherence signals.
The transverse bunch decoherence and the resulting coherent spectra are measured
in the SIS18 synchrotron at GSI Darmstadt for different bunch parameters.
Particle tracking simulations together with an analytical model
are used to describe the modifications in the 
decoherence signals and in the coherent spectra
due to space charge and the rf bucket nonlinearity.
\end{abstract}

\maketitle

\section{INTRODUCTION}

Transverse coherent oscillations of bunches induced by a fast kicker magnet
are routinely used in synchrotron or storage ring to measure for example the tune or other
ring parameters, see e.g.\ \cite{jones06}. 
The transverse offset of a bunch, averaged over the bunch length, can be
recorded every single turn. The spectrum is then
concentrated around the base-band $Q_{f0} f_0$,
where $Q_{f0}$ is the fractional part of the betatron tune $Q_0$
and $f_0$ is the revolution frequency.
This diagnostics is usually used for time-resolved and
very accurate measurements of the tune $Q_{f0}$.

Transverse bunch decoherence is a process of
a turn-to-turn reduction of the total bunch offset signal
after an initial bunch displacement.
In a linear focusing lattice
the bunch decoherence is a manifestation of the lattice chromaticity $\xi$
where the synchrotron dynamics also plays an important role,
causing the signal recoherence exactly after the synchrotron period.
Other damping mechanisms, as due to lattice nonlinearities,
additionally damp the transverse oscillations.
Transverse decoherence is often used as a machine diagnostics tool.
Undesired transverse bunch oscillations can also appear
after the bunch-to-bucket transfer between synchrotrons.
In order to use transverse decoherence as a diagnostics tool for intense bunches of arbitrary length
and also to control undesired oscillations of such bunches it is important to understand 
the decoherence in the presence of transverse space charge and nonlinear synchrotron oscillations.  

We demonstrate that the decoherence signal 
can be explained in terms of the transverse head-tail bunch mode spectrum.
For finite chromaticity also the $k>0$ head-tail modes contribute 
to the bunch coherent spectrum.
The shift of the head-tail mode frequencies due to space charge and wall currents can be well
explained in terms of the analytical expressions for an airbag bunch distribution
\cite{blask98,boine2009}.
The head-tail mode frequencies are also modified
by changes in the individual particle synchrotron frequency.
In long bunches one has to account for the spread of the synchrotron frequencies.
Both, transverse space charge and nonlinear synchrotron oscillations
are important to understand the decoherence signals and transverse spectra.
We demonstrate that, once the spectrum- and decoherence modifications are understood,
they can be used to extract useful information about the bunches.

In this work we describe measurements of transverse bunch spectra
and decoherence signals obtained in the heavy-ion synchrotron SIS18 at GSI Darmstadt.
The observed modification of the head-tail spectrum and of the decoherence signal 
caused by transverse space charge and nonlinear synchrotron oscillations
are explained in terms of our theoretical approach.
This approach is based on an expansion of an analytical theory for
head-tail modes in combination with particle tracking simulations.

In Sec. \ref{sec:theory} we use theoretical and numerical approaches to
analyse the effects of space charge and nonlinear synchrotron motion on the transverse spectra
and on the bunch decoherence signal.
We show that a simple model for the head-tail mode frequencies 
with fitting parameters can be used
to explain the numerically obtained spectrum modifications
as well as the bunch decoherence as a function of the chromaticity.
In Sec.\,III the results of measurements performed at the SIS18 synchrotron are presented.
The space charge tune shifts determined from the transverse spectra are summarized,
the role of nonlinear synchrotron motion is demonstrated and transverse bunch 
decoherence signals measured for different bunch conditions
are presented and explained. The work is concluded in Sec.\,IV.

\section{THEORY AND NUMERICAL SIMULATIONS}\label{sec:theory}

The Fourier transformation of the transverse
bunch signal provides peaks at frequencies
which represent the bunch eigenmodes, also called head-tail modes.
For short, low intensity bunches (the synchrotron frequency $f_s=Q_s f_0$ does
not depend on the amplitude, no collective effects)
the transverse spectrum has peaks at
$\Delta Q = Q-Q_{f0} = 0$ for $k=0$, $\Delta Q = \pm Q_s$ for $k = \pm 1$,
$\Delta Q = \pm 2 Q_s$ for $k = \pm 2$, and so forth.
Collective effects, like transverse space charge or
ring impedances change the bunch eigenfrequencies
and thus shift the peaks in the transverse spectrum.

Transverse space-charge effects are described 
by the characteristic tune shift,
\begin{eqnarray}
\Delta Q_{\rm sc} =
\frac{\lambda_0 r_p R}{\gamma^3 \beta^2 \varepsilon_\perp} \ ,
\label{eq09}
\end{eqnarray}
where $R$ is the ring radius,
$\beta$ and $\gamma$ are the relativistic parameters,
$r_p=q_{\rm ion}^2/(4 \pi \epsilon_0 m c^2)$ is the classical particle radius,
$\lambda_0$ is the peak line density (at the bunch center),
and $\varepsilon_\perp$ is the transverse total emittance.
This tune shift corresponds to a round cross-section
with a transverse K-V distribution
and is defined as the modulus of the negative shift.
In a rms-equivalent bunch with the Gaussian transverse profile,
i.e.\ the transverse rms emittance is $\varepsilon_x=\varepsilon_\perp/4$,
the maximum space-charge tune shift is twice of this value,
$\Delta Q_{\rm sc}^{\rm max} = 2 \Delta Q_{\rm sc}$.
In the case of an elliptic transverse cross-section
with the rms emittances $\varepsilon_y, \varepsilon_x$,
the parameter $\varepsilon_\perp$ in Eq.\,(\ref{eq09})
should be replaced by
\begin{eqnarray}
\varepsilon_\perp =
2 \Bigl( \varepsilon_y + \sqrt{\varepsilon_y\varepsilon_x
\frac{Q_{0y}}{Q_{0x}} } \Bigr) \ ,
\label{eq10}
\end{eqnarray}
here for the vertical ($y$) plane, for the horizontal plane correspondingly.
The parameter for the effect of space charge in a bunch
is defined as a ratio of the characteristic space-charge tune shift Eq.\,(\ref{eq09})
to the small-amplitude synchrotron tune,
\begin{eqnarray}
q = \frac{\Delta Q_{\rm sc}}{Q_{s0}} \ .
\label{eq11}
\end{eqnarray}

\subsection{LONGITUDINAL DIPOLE FREQUENCY}

\begin{figure}[h!]
\includegraphics*[width=.5\linewidth]{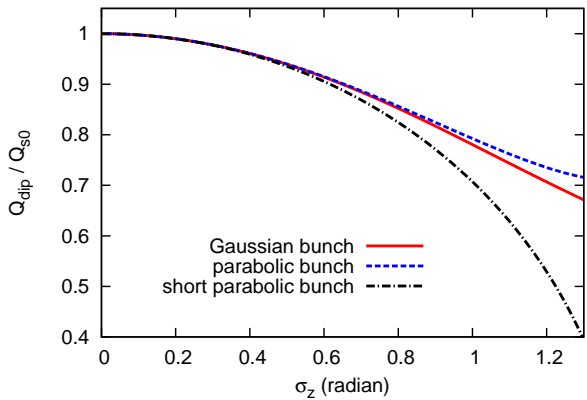}
\caption{\label{fg05}
The longitudinal dipole oscillation frequency
as a function of the rms bunch length.
The red curve is obtained using Eq.\,(\ref{eq01}) for a Gaussian bunch,
the blue dashed curve is given by Eq.\,(\ref{eq02})
and the back chain curve is given by Eq.\,(\ref{eq03}).
}
\end{figure}

An important parameter for head-tail bunch oscillations
in long bunches is the effective synchrotron frequency
which will be different from the small-amplitude synchrotron frequency 
in short bunches. We will show
that in long bunches the longitudinal dipole frequency $Q_{\rm dip} f_0$ can be chosen
as a substitute for the small-amplitude incoherent synchrotron frequency $Q_{s0} f_0$.
The longitudinal dipole frequency can be accurately measured from the bunch signal, 
as we will show in the experimental part of this paper.
The frequency of small-amplitude dipole oscillations
can be calculated as \cite{boine_rf2005}
\begin{eqnarray}
\frac{Q_{\rm dip}^2}{Q_{s0}^2} =
2 \int_0^{\tau_{\rm max}} \frac{V_{\rm rf}}{V_0} \lambda^\prime (\tau) d \tau \ ,
\label{eq01}
\end{eqnarray}
where the rf voltage form is $V_{\rm rf} = V_0 \sin (\tau)$,
$\tau$ is rf bucket radian and $\lambda$ is the line density.
The small-amplitude bare synchrotron tune is given by
\begin{eqnarray}
Q_{s0}^2 =
\frac{q_{\rm ion} V_0 h |\eta|}{2 \pi m \gamma \beta^2 c^2} \ ,
\label{eq07}
\end{eqnarray}
where $\eta$ is the machine slip factor
and $h$ is the rf harmonic number.
The dependence of $Q_{\rm dip}$ on
the rms bunch length $\sigma_z$
for a Gaussian bunch is shown (red curve) in Fig.\,\ref{fg05}.
The bunch length $\sigma_z$ is dimensioned in radian
of the rf bucket, i.e.\ $\sigma_z = L_{\rm rms} h/R$,
where $L_{\rm rms}$ is the rms bunch length in meter.

For a parabolic longitudinal distribution (or elliptic bunch)
with the total half-length $\tau_p = \sqrt{5} \sigma_z$ one obtains
the analytic expression \cite{boine_rf2005},
\begin{eqnarray}
\frac{Q_{\rm dip}^2}{Q_{s0}^2} =
\frac{2 \tau_p - \sin (2 \tau_p)}{4 \sin (\tau_p) -
4 \tau_p \cos (\tau_p)} \ ,
\label{eq02}
\end{eqnarray}
which can be approximated in the case of a short bunch as
\begin{eqnarray}
\frac{Q_{\rm dip}}{Q_{s0}} =
\sqrt{1 - \frac{\sigma_z^2}{2}} \ .
\label{eq03}
\end{eqnarray}
From the comparison in Fig.\,\ref{fg05} it follows that
for short bunches with $\sigma_z \lesssim 0.6$
the approximation Eq.\,(\ref{eq03}) is sufficient.
For long bunches with $\sigma_z \gtrsim 1$ the
dipole frequencies for Gaussian and parabolic bunches start to differ.

\subsection{SPECTRUM OF A LONG BUNCH WITH SPACE CHARGE}

We use particle tracking simulations \cite{boine2006, rumolo2002}
in order to investigate the combined effect of space charge and
nonlinear synchrotron motion on transverse head-tail oscillations.
The numerical codes have been validated \cite{kornilov-icap09} using
analytic results \cite{blask98}. For the transverse space charge force,
a frozen electric field model is used,
i.e.\ a fixed potential configuration which follows
the center of mass for each bunch slice.
This approach is justified in the rigid-slice regime
and can be considered as a reasonable approach for
moderate and strong space charge \cite{burov-lebed2009, burov2009}.
A round transverse cross-section and a Gaussian transverse beam profile
were used in the simulations in this work.

\begin{figure}[h!]
\includegraphics*[width=.5\linewidth]{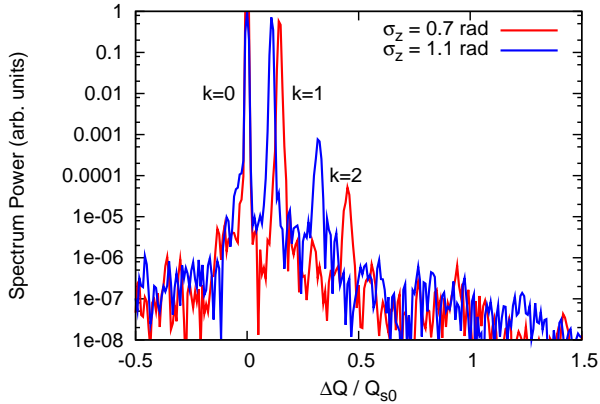}
\caption{\label{fg01}
Example transverse spectra
of long bunches from particle tracking simulations,
with space charge and nonlinear synchrotron motion taken into account.
Bunches with two different rms length $\sigma_z$ are assumed,
the space charge parameter $q=8$ and
the bare synchrotron tune $Q_{s0}$ are kept constant.
The spectra clearly show the head-tail modes
$k=0$, $k=1$ and $k=2$.
The tune shift $\Delta Q$ is related to the
bare betatron tune as $\Delta Q = Q-Q_{f0}$.
}
\end{figure}

\begin{figure}[h!]
\includegraphics*[width=.5\linewidth]{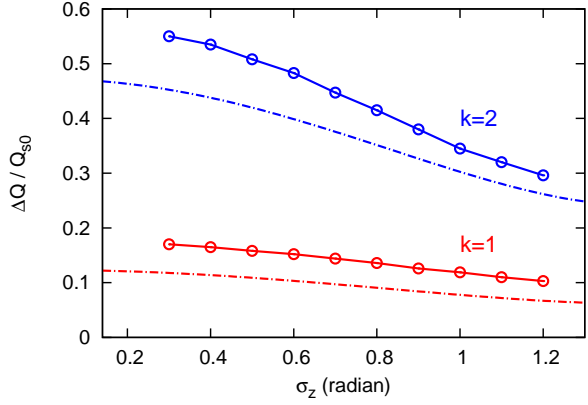}
\caption{\label{fg04}
Results of a simulation scan (circles) over the rms bunch length
for a bunch with space charge parameter $q=8$.
Red corresponds to the $k=1$ head-tail mode and 
blue to $k=2$ modes.
For comparison, the chain curves show an estimation
using Eq.\,(\ref{eq05}) with $q_*=Q_{\rm dip}/Q_{s0}$.
}
\end{figure}

\begin{figure}[h!]
\includegraphics*[width=.5\linewidth]{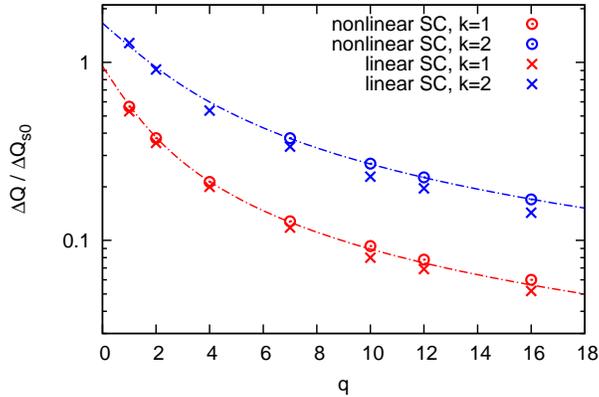}
\caption{\label{fg02}
Results of a simulation scan over the space charge parameter
for a bunch with the rms length $\sigma_z=1.06$\,rad.
The crosses show the eigenfrequencies of the modes
$k=1$ and $k=2$ for bunches with a transverse K-V distribution,
while the circles are for bunches with a Gaussian transverse
profile.
The chain curves are given by Eq.\,(\ref{eq05})
with the coefficients $q_*=0.95$ for $k=1$
and $q_*=0.83$ for $k=2$,
which corresponds to the results summarized in Fig.\,\ref{fg03}.}
\end{figure}

Figure\,\ref{fg01} demonstrates differences in the
transverse mode frequencies for bunches of different lengths,
and with all the other parameters kept identical, including
the space charge parameter $q=8$.
The three lowest-order modes can be seen very well, the modes of the longer bunch
are shifted closer to the bare betatron tune than those of the shorter bunch.
In order to describe the bunch spectrum for
arbitrary bunch length and space charge strength,
simulation scans over different parameters have been performed.
Our simulation results suggest that the airbag bunch model \cite{blask98}
can be applied to the head-tail modes in a long Gaussian bunch,
\begin{eqnarray}
\frac{\Delta Q_k}{Q_{s0}} =
-\frac{q}{2} \pm \sqrt{\frac{q^2}{4} + k^2 q_*^2} \ ,
\label{eq05}
\end{eqnarray}
where $q_*=Q_{s*}/Q_{s0}$ is a characteristic parameter depending on the bunch length 
and the nonlinear synchrotron oscillations. In our case $q_*$ is used as a fitting parameter.
Keeping the space charge parameter constant,
the bunch length has been varied and the resulting
eigenfrequencies analyzed, see Fig.\,\ref{fg04}
for a scan with $q=8$.
We observe substantial changes in the bunch mode frequencies
with increasing bunch length.
The parameter $q_*$ has been obtained from these simulation scans.
Figure\,\ref{fg02} shows a comparison between simulation results
and the model Eq.\,(\ref{eq05}) for a fixed bunch length
and for different space charge parameters.
The plot demonstrates that the model Eq.\,(\ref{eq05}) is fairly
accurate over the parameter range of the interest.
As we additionally show in Fig.\,\ref{fg02},
there is a small difference between transverse Gaussian bunch profiles 
(with nonlinear transverse space charge)
and transverse K-V distributions (with linear space charge).
In our simulations we use the more realistic
Gaussian profile.

The chain curves in Fig.\,\ref{fg04} show that it would be
not correct to use the longitudinal dipole tune $Q_{\rm dip}$
for the parameter $Q_{s*}$. 
An interesting observation is that
the type of the dependence of the mode frequencies on 
the bunch length is similar to $Q_{\rm dip}$,
being however slightly different.
Also, the scale factor between $Q_{\rm dip}$ and the real $\Delta Q$
is quite different for $k=1$ and $k=2$.
The bare synchrotron tune, which would mean $q_*=1$, is not
an adequate value, too, $\Delta Q$ would then be a constant for changing
bunch length and it would correspond to the value
of the chain curve at small $\sigma_z$.

\begin{figure}[h!]
\includegraphics*[width=.5\linewidth]{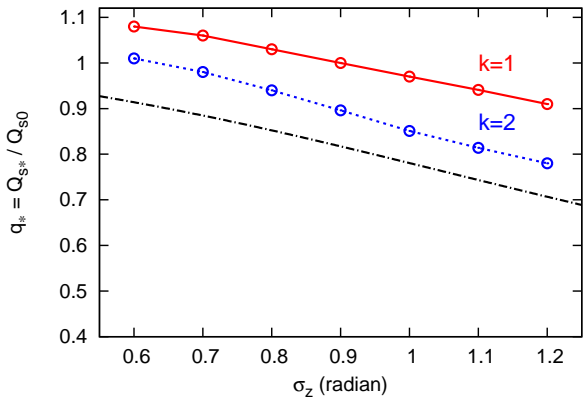}
\caption{\label{fg03}
Summary of the simulation scans for the effect of the bunch length
on the eigenfrequencies of the head-tail modes $k=1$ and $k=2$
with space charge.
For comparison, the chain curve shows the longitudinal
dipole frequency from Eq.\,(\ref{eq01}) for a Gaussian bunch.
}
\end{figure}

Simulation results for practical usage are presented in Fig.\,\ref{fg03}.
These $q_*$ values can be included in Eq.\,(\ref{eq05})
in order to estimate the space charge tune shift
of the bunch eigenfrequencies for a given bunch length.
The chain line demonstrates again the difference between
$Q_{s*}$ which describes the tune shift and the longitudinal dipole
frequency.

\subsection{TRANSVERSE DECOHERENCE}

\subsubsection{LINEAR DECOHERENCE}
\begin{figure}[h!]
\includegraphics*[width=.6\linewidth]{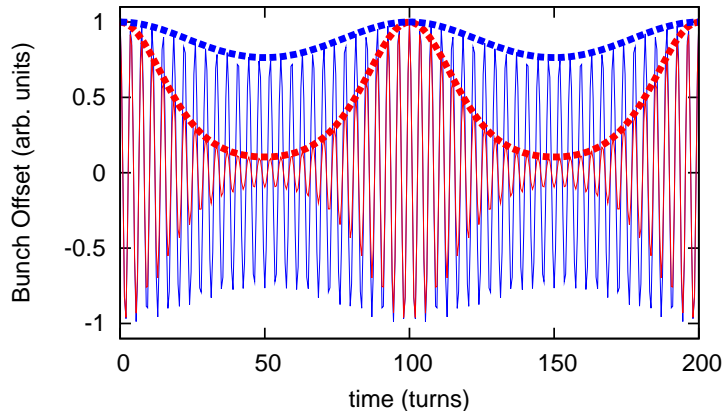}
\caption{\label{fg08}
A particle tracking simulation for a Gaussian bunch
after an offset kick $\overline{x}(\tau)={\rm const}$
without space charge and for a linear rf bucket, $Q_s=Q_{s0}=0.01$.
The full lines show the time evolution of the bunch offset
for the chromaticities
$\xi Q_0=-4.3$ (blue) and $\xi Q_0=-12.5$ (red),
the dashed lines are analytical results and are given by Eq.\,(\ref{eq04}).
}
\end{figure}

First, we discuss the linear transverse decoherence
due to chromaticity,
i.e.\ the only source of the tune shift is the linear
dependence of the betatron tune shift on the momentum shift
$\Delta Q_\xi / Q = \xi \Delta p / p$.
As a result of an initial transverse displacement
$\overline{x}(\tau)=A_0$,
a bunch oscillates in the corresponding plane (here $x$).
As we consider the linear case, all the particles
have the identical synchrotron frequency $Q_s f_0$.
The betatron phase shift related to the
bare tune $Q_0$ has a harmonic dependency along a synchrotron period.
Hence, after one synchrotron oscillation,
the betatron phase shift is exactly compensated
and the transverse amplitude is equal to the
initial displacement $A_0$.
Assuming the Gaussian momentum distribution,
the amplitude of the bunch offset evolves
with the turn number $N$ as \cite{meller87}
\begin{eqnarray}
A(N) = A_0 \exp \Biggl\{ -2
\Bigl( \frac{\xi Q_0 \delta_p}{Q_s}
\sin (\pi Q_s N)
\Bigr)^2 \Biggr\} \ ,
\label{eq04}
\end{eqnarray}
here
$\delta_p$ is the normalized rms momentum spread.
Figure\,\ref{fg08} shows an example for bunch decoherence
after a rigid kick.
It demonstrates that a higher chromaticity
provides a faster decoherence, and that after the synchrotron period
$N_{\rm s}=1/Q_{\rm s}$ the initial offset amplitude appears again,
which is called recoherence.

\subsubsection{DECOHERENCE WITH SPACE CHARGE}

\begin{figure}[h!]
\includegraphics*[width=.6\linewidth]{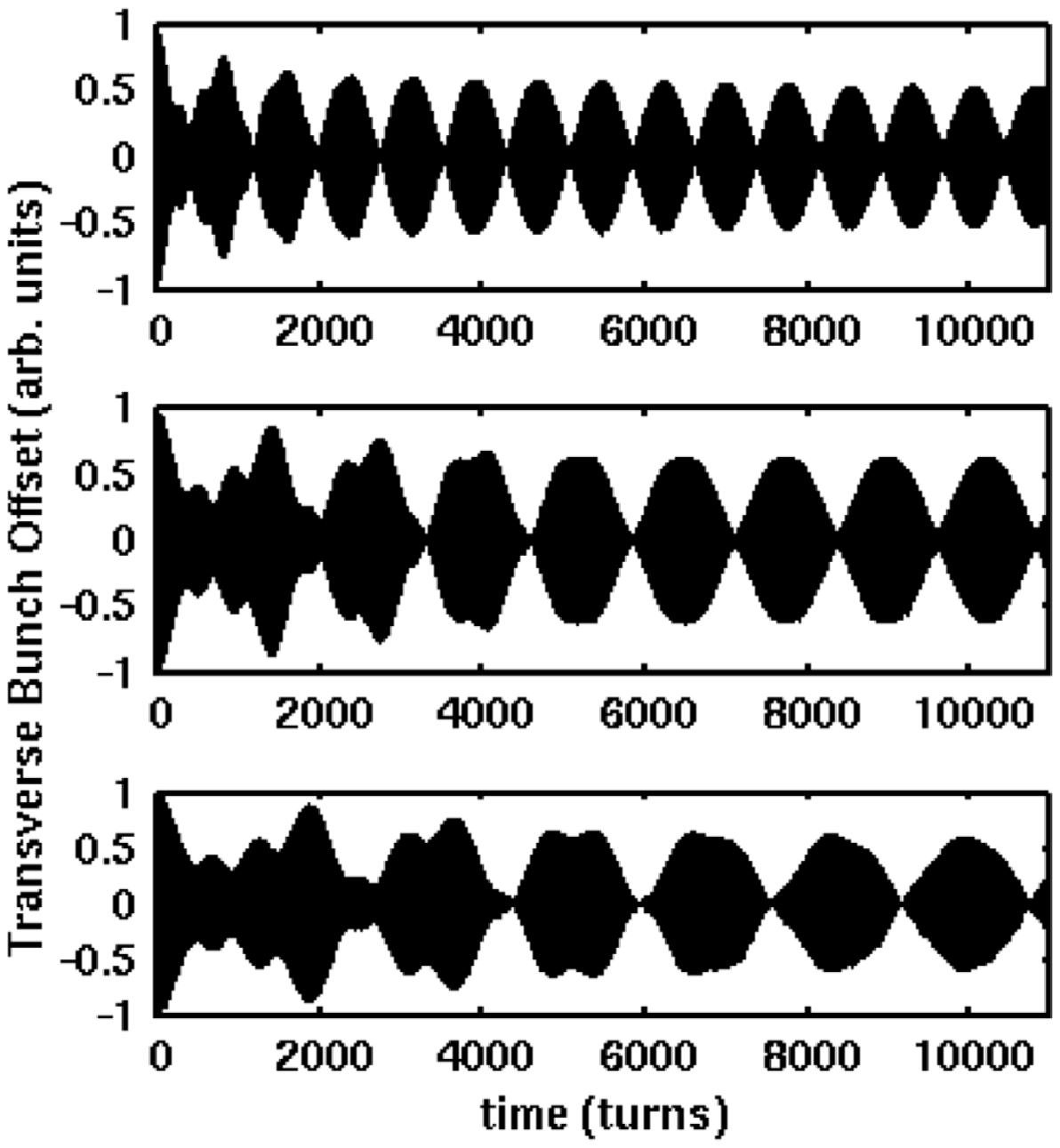}
\caption{\label{fg06}
Transverse bunch decoherence from
particle tracking simulations for a Gaussian bunch
after a rigid kick $\overline{x} (\tau) = {\rm const}$
for different space charge parameters.
Top plot: $q=7$, middle plot: $q=12$
and for the bottom plot $q=16$.
The bare synchrotron tune is $Q_{s 0} = 0.01$,
i.e.\ the low-intensity recoherence has a periodicity of 100\,turns.
After the transition period of higher-order mode damping,
the periodicity always corresponds
to the frequency difference
$\Delta Q = Q_{k=1} - Q_{k=0}$.
Top plot: $\Delta Q_{k=1} = 0.13 \ Q_{s0}$ (periodicity 770 turns),
middle plot: $\Delta Q_{k=1} = 0.079 \ Q_{s0}$ (periodicity 1270 turns),
and for the bottom plot
$\Delta Q_{k=1} = 0.061 \ Q_{s0}$ (periodicity 1640 turns).
}
\end{figure}

Transverse space charge causes a betatron frequency shift,
which depends on the particle transverse amplitude
and on the longitudinal particle position in the bunch.
The decoherence behaviour is thus very different from the
linear decoherence at low bunch intensities Eq.\,(\ref{eq04}).
Figure\,\ref{fg06} shows examples of the bunch oscillations
after a rigid kick for three different values
of the space-charge parameter.
The chromaticity corresponds to $\chi_b=4.5$,
where $\chi_b = Q_0 \xi L_b / (\eta R)$ is
the chromatic phase shift over the bunch length,
the bunch rms length is $\sigma_z=1.06$\,rad.
We observe the periodic recoherence
with the periodicity 770 turns ($q=7$, top), 1270 turns ($q=12$, middle)
and 1640 turns ($q=16$, bottom), while the low-intensity recoherence
would have a periodicity of 100 turns for the same parameters.

The key in understanding of the decoherence
for a bunch with transverse space charge is the representation
of the initial kick as a superposition of the
bunch head-tail eigenmodes,
\begin{eqnarray}
A_0 = \sum_k a_k \exp
\Bigl(-i \frac{\chi_b \tau}{\tau_b} + i \phi_k \Bigr)
\overline{x}_k (\tau) \ ,
\label{eq08}
\end{eqnarray}
where we have extracted the chromatic phase shift along the bunch
with the corresponding phase $\phi_k$ for each eigenfunction.
The second key is the fact that the different eigenmodes
are prone to Landau damping mechanisms, but with different intensity thresholds and damping rates.
Landau damping due to the space charge tune spread along the bunch
\cite{burov2009, balb2009, kornilov_prstab10} is the most important 
mechanism in the beam parameter regime considered in the simulations of this work.
In the presence of space charge especially the negative 
and the high-$k$ eigenmodes present in the initial kick Eq.\,(\ref{eq08})
are quickly suppressed, so that after a transition period
a mixture of the survived eigenmodes continues to oscillate.

In Ref.\,\cite{kornilov_hb10d} we have discussed in detail the case
$q=1$, where all the head-tail modes $k \geq 1$ are strongly
suppressed by Landau damping such that the mode $k=0$ is left alone.
For stronger space charge, as in Fig.\,\ref{fg06},
the modes $k \geq 2$ are damped and the resulting oscillation
is the mixture of the $k=0$ and $k=1$ modes.
The recoherence periodicity seen in Fig.\,\ref{fg06} corresponds exactly
to the frequency difference between these two modes,
as it is the case for the wave beating.
In a real machine there are often nonlinear damping mechanisms which would further suppress
the $k=0$ and $k=1$ modes, but in the simulation
we only have the space charge induced Landau damping which is zero for the $k=0$ mode
and is very weak for the $k=1$ mode at these $q$ parameters.

\begin{figure}[h!]
\includegraphics*[width=.6\linewidth]{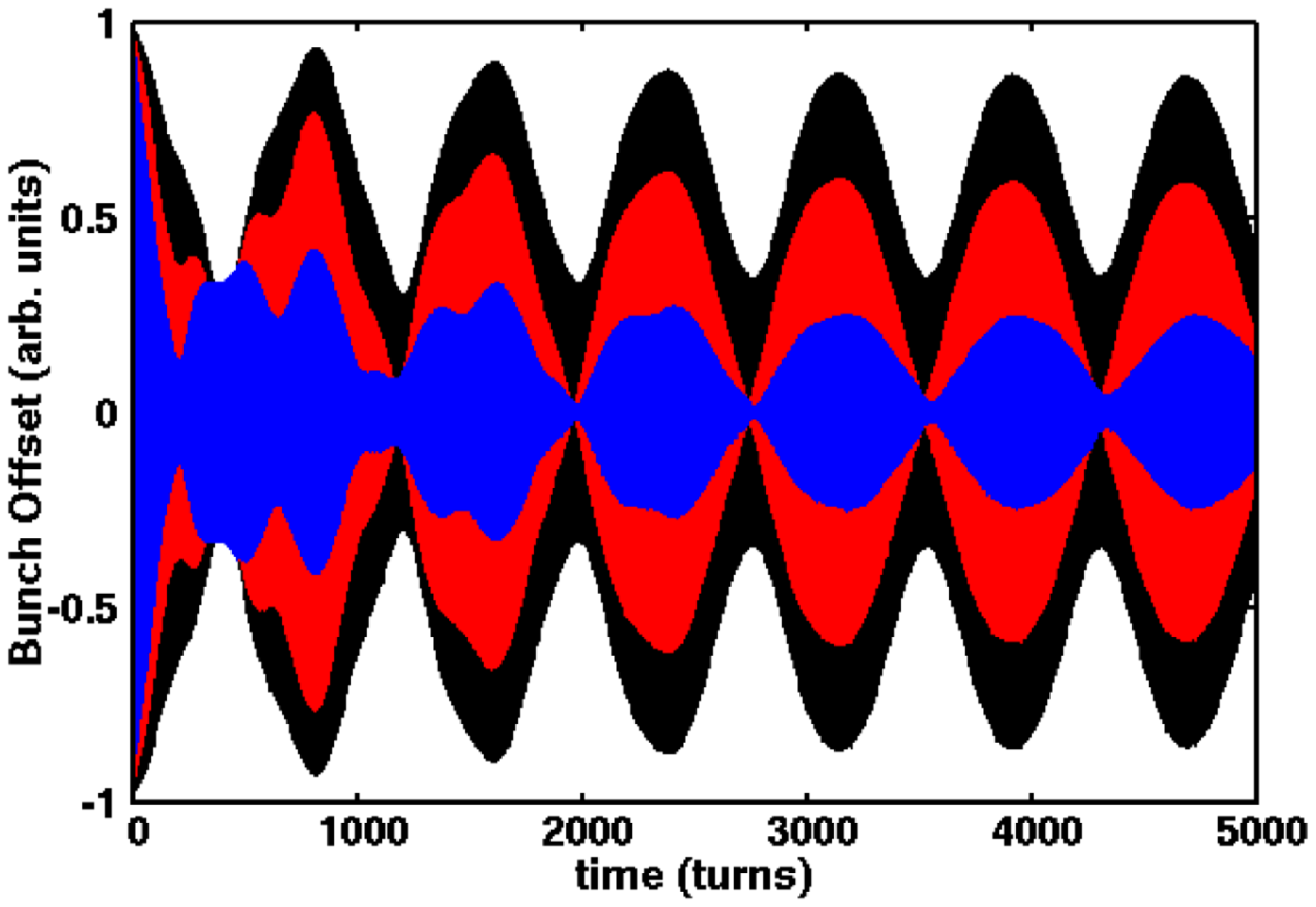}
\caption{\label{fg07}
 Transverse bunch decoherence for a bunch with space charge parameter $q=7$
from particle tracking simulation for different chromaticities.
The black curve:
$\chi_b=3$,
the red curve:
$\chi_b=4.5$
and for the blue curve
$\chi_b=6$.
The recoherence results from a mixture of the $k=0$ mode
and $k=1$ mode, $\Delta Q_{k=1} = 0.13 \ Q_{s0}$ (periodicity 770 turns).
}
\end{figure}

It is obvious, and can be seen in Eq.\,(\ref{eq08}),
that the composition of the eigenmodes after a rigid kick depends on the chromaticity.
This is also demonstrated in Fig.\,\ref{fg07} which shows 
a comparison of the bunch decoherence for three different chromaticities.
The bunch parameters correspond to Fig.\,\ref{fg06}, the
space charge parameter is chosen as $q=7$.
We see that the periodicity of 770 turns does not change. It 
corresponds to the frequency difference $\Delta Q = Q_{k=1} - Q_{k=0}=0.13 Q_{s0}$.
The reason for the different oscillation amplitudes in Fig.\,\ref{fg06}
is the increasing contribution of higher-order modes $k \ge 2$
with growing $\xi$ in the eigenmode mixture of the initial rigid bunch offset (see Eq.\,\ref{eq08}).
Recall that these modes are quickly suppressed for the parameters of the bunch and
the resulting recoherence is a beating of the remaining $k=0$ and $k=1$ modes.

\begin{figure}[h!]
\includegraphics*[width=.6\linewidth]{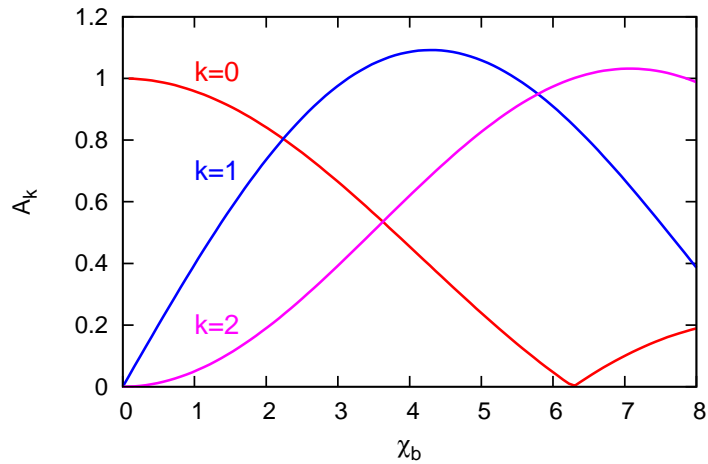}
\caption{\label{fg09}
Relative amplitudes of the airbag-bunch eigenmodes
for the rigid offset $\overline{x} (\tau)={\rm const}$
as a function of the chromatic phase shift
$\chi_b=Q_0 \xi L_b / (\eta R)$.
}
\end{figure}

The airbag \cite{blask98} eigenmodes $\overline{x}_k (\tau)=
A \cos (k \pi \tau / \tau_b)$
can be taken as a reasonable approximation \cite{kornilov_prstab10}
of the eigenfunctions in a Gaussian bunch.
The rigid offset decomposition Eq.\,(\ref{eq08}) can then be solved
and the resulting mode coefficients are
$a_0=2 / \chi_b \sin (\chi_b/2)$,
$a_1=4 \chi_b/ |\chi_b^2 - \pi^2| \cos (\chi_b/2)$,
$a_2=4 \chi_b/ |\chi_b^2 - 4 \pi^2| \sin (\chi_b/2)$.
The negative modes have the same coefficients
but can be disregarded in the case of a bunch with space charge \cite{burov2009, kornilov_prstab10}, 
because of their large damping rates. 
These coefficients are plotted in Fig.\,\ref{fg09},
where we see that for the chromaticity range of interest
the relative part of the $k=2$ mode increases with growing $\chi_b$.
The higher-order modes follow this trend.
The contribution of the $k=0$ and $k=1$ modes thus decreases
as we also can observe in the simulations, see Fig.\,\ref{fg07}.
A perfect agreement with the coefficients in Fig.\,\ref{fg09}
can not be expected, since the analytical model is for an
airbag \cite{blask98} bunch.

\clearpage
\section{MEASUREMENTS}

Transverse decoherence experiments have been performed
in the heavy ion synchrotron SIS18 \cite{sis18} at GSI Darmstadt.
Bunches of Ar$^{18+}_{40}$ ions were stored at the energy of 100\,MeV/u
and kicked transversally with a kick duration of one turn.
The rf harmonic number was $h=4$ and all the four bunches
had generally an identical behaviour.
The Beam Position Monitors (BPMs) provide a higher quality signal in the vertical plane
than in the horizontal one due to a smaller plate gap,
thus we use the vertical BPM signals in the results presented here.
The vertical bare tune was around $Q_0=4.31$ although
it could vary for different intensities and machine parameters.
SIS18 general parameters are: $R=34.492$\,m, $\gamma_t=5.45$,
$\xi \approx -1.4$.

\begin{figure}[h!]
\includegraphics*[width=.7\linewidth]{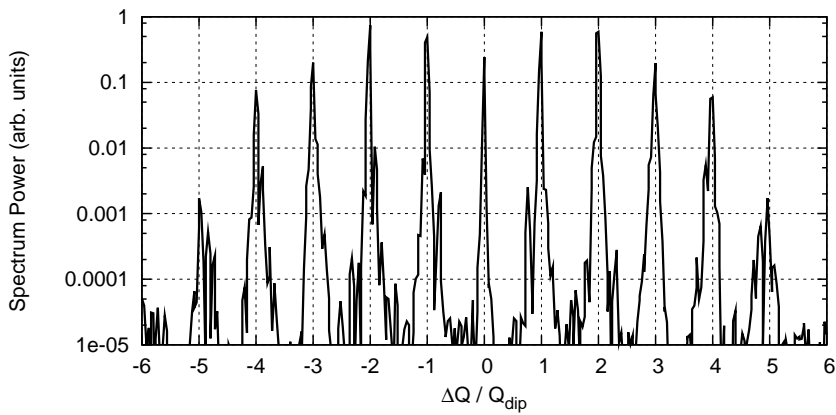}
\caption{\label{fg20}
An example for a longitudinal
spectrum from BPM bunch measurements at SIS18.
The spectrum corresponds to the 22nd harmonic
at the frequency of 13.073\,MHz,
$\Delta Q = (f - n f_0)/f_0$.
}
\end{figure}

Similar to the theory section, first we discuss
the longitudinal dipole frequency.
Figure\,\ref{fg20} demonstrates the bunch spectrum
obtained from the sum BPM signal.
The satellites of the central frequency are well resolved,
the peaks are equidistant what provides the longitudinal dipole frequency.
The longitudinal dipole frequency determined
by this way is $Q_{\rm dip}=2.5\times10^{-3}$,
the peak rf voltage was $V_0=9$\,kV here.
The bare synchrotron tune
can also be accurately determined using Eq.\,(\ref{eq07}) and it is 
$Q_{s0}=3.24\times10^{-3}$ in this case.
Note the large difference between the bare synchrotron
frequency and the dipole frequency.
Using the curves from Fig.\,\ref{fg05} we can obtain
the rms bunch length
$\sigma_z=1.0$\,rad,
which is a typical length in the experiments at SIS18.

\begin{figure}[h!]
\includegraphics*[width=.7\linewidth]{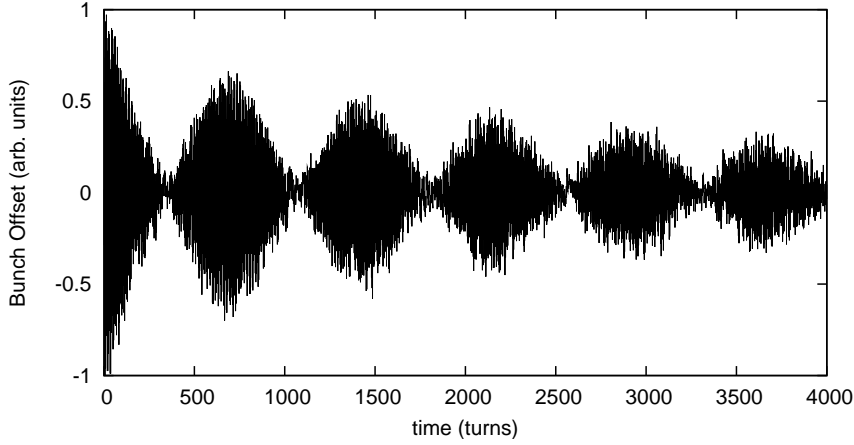}
\caption{\label{fg21}
Time evolution of the bunch offset in the vertical plane
at SIS18 after a transverse kick.
The recoherence periodicity corresponds to the mix of the dominating head-tail modes
$k=0$ and $k=2$ with $\Delta Q_{k=2}=1.35\times10^{-3}$,
giving the periodicity of 740\,turns.
}
\end{figure}

\begin{figure}[h!]
\includegraphics*[width=.7\linewidth]{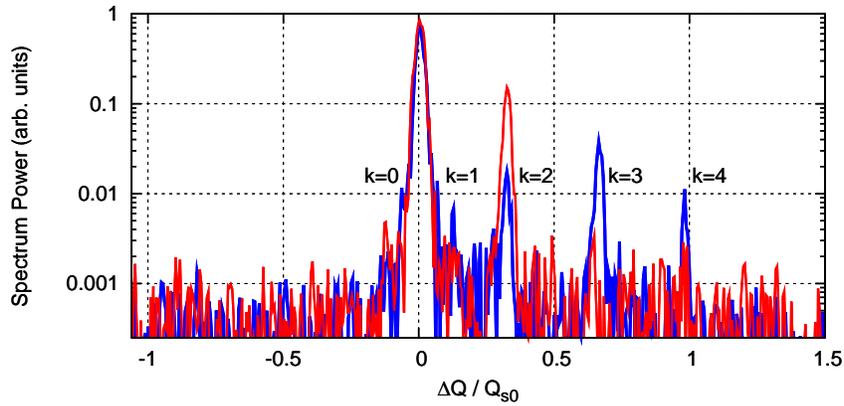}
\caption{\label{fg22}
Transverse coherent spectrum
for the bunch from Fig.\,\ref{fg21}, $Q_{s0}=4.04\times10^{-3}$.
The red spectrum is obtained from a frequency analysis
of the complete bunch offset,
while the blue spectrum is a result of a frequency analysis
for the bunch head.
In the complete-bunch spectrum the mode $k=2$ dominates,
and the bunch head spectrum reveals the uneven modes
$k=1$, $k=3$ but also the mode $k=4$.
The frequencies of the head-tail modes
provide the space charge parameter $q \approx 10$,
see blue circles in Fig.\,\ref{fg30}.
}
\end{figure}

\begin{figure}[h!]
\includegraphics*[width=.5\linewidth]{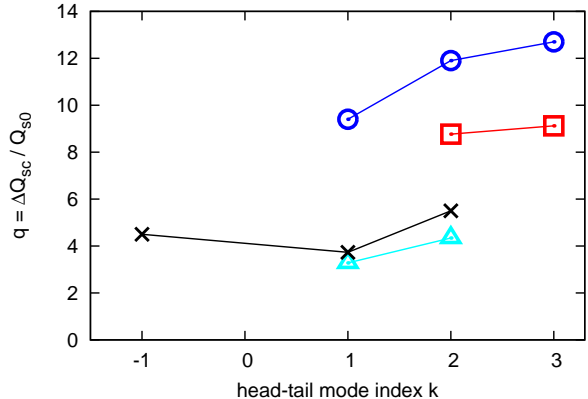}
\caption{\label{fg30}
Summary for the space charge parameter determined
from the coherent head-tail spectra of different Ar$^{18+}_{40}$ bunched beams
in the SIS18 synchrotron.
The method is given by Eq.\,(\ref{eq06}),
with the coefficients $q_*$ corresponding to Fig.\,\ref{fg03}.
The spectra are shown in Fig.\,(\ref{fg22}) (blue circles),
Fig.\,\ref{fg24} (red squares),
Fig.\,\ref{fg26} (black crosses) and
in Fig.\,\ref{fg29} (cyan triangles).
}
\end{figure}

The first example for the decoherence measurements
is presented in Fig.\,\ref{fg21} and Fig.\,\ref{fg22}.
Figure\,\ref{fg21} shows the turn-per-turn transverse
bunch offset after the kick.
Figure\,\ref{fg22} demonstrates the spectrum of these
bunch oscillations, the frequency on the horizontal axis
is normalized by the bare synchrotron tune.
The red line is for the spectrum of the whole bunch
and shows mainly peaks of two modes which we can identify
as the $k=0$ mode and the $k=2$ mode.
If we calculate a Fourier transform for the bunch head,
its spectrum (the blue line) clearly reveals other peaks,
so that we can identify five head-tail modes, see Fig.\,\ref{fg22}.
The spectrum is very different from the case without collective effects:
the lines are not equidistant, the negative-modes ($k<0$) are suppressed.
The fact that the mode tune shifts are consistent with the space-charge model
can be proved by calculating the space charge parameter,
\begin{eqnarray}
q = \frac{k^2 q_*^2 - ( \Delta Q_k / Q_{s0} )^2 }{
\Delta Q_k / Q_{s0} } \ ,
\label{eq06}
\end{eqnarray}
which corresponds to the model Eq.\,(\ref{eq05}).
The synchrotron oscillation parameter $q_*$
for the modes $k=1$ and $k=2$ is obtained from
the results given in Fig.\,\ref{fg03}.
$\Delta Q_k$ is the tune shift of the bunch mode from the measured spectrum.
Here and for the examples to follow we summarize the space charge parameters $q$
obtained from the different eigenfrequencies of the spectra in Fig.\,\ref{fg30}.
The relevant bunch parameters are summarized in Table\,1.
The values for the modes from Fig.\,\ref{fg22}
are shown in Fig.\,\ref{fg30} with the blue circles, $q \approx 10$.
Since this was a rather short bunch, $\sigma_z=0.66$,
the $q_*$-parameter was close to 1.0
and thus it was possible
to estimate the space charge parameter for the $k=3$ mode as well.

\begin{table}[h!]
\begin{center}
\begin{tabular}{|c|c|c|c|c|c|c|}
\hline
signals & symbols in Fig.\,\ref{fg30} & $\sigma_z$ & $\chi_b$ &
$Q_{\rm dip}, 10^{-3}$ & $Q_{s0}, 10^{-3}$ &  $q$ \\
\hline
Figs.\,\ref{fg21},\,\ref{fg22} & blue circles &
\ \ 0.66 \ \ & \ $\approx$ 3 \ & 3.63 & 4.0 &  $\approx$ 10 \\
\hline
Figs.\,\ref{fg23},\,\ref{fg24} & red squares &
1.15 & $\approx$ 5 & 2.35 & 3.24 & $\approx$ 9 \\
\hline
Figs.\,\ref{fg25},\,\ref{fg26} & black crosses &
1.2 & $\approx$ 5 & 2.28 & 3.24 & $\approx$ 4.5 \\
\hline
Figs.\,\ref{fg28},\,\ref{fg29} & cyan triangles &
1.0 & $\approx$ 0 & 2.5 & 3.24 & $\approx$ 4 \\
\hline
\end{tabular}
\end{center}
\caption{Bunch parameters for the signals shown in this paper
and the space-charge parameter $q$ obtained from the transverse spectra.
The $q$--values from the different head-tail modes for every bunch
are summarized in Fig.\,\ref{fg30}.
In the first case the rf voltage was $V_0=14$\,kV,
in the last three cases $V_0=9$\,kV.
}
\end{table}

Figure\,\ref{fg30} demonstrates a certain consistency
between different head-tail modes for the space charge parameter,
that, however, can not be expected perfect.
The model Eq.\,(\ref{eq05}) is based on the airbag \cite{blask98} bunch
which is a reasonable,
but still an approximation for a Gaussian bunch \cite{kornilov_prstab10}.
The bunch spectra are also weakly affected by the facility impedances,
image charges and nonlinear field components neglected in our analysis.
Finally, in our simulations Gaussian bunch profiles
in the transverse and in the longitudinal plane
have been assumed. It is a good, but not exact description
for the bunches in the machine experiments.

The space-charge parameter $q=\Delta Q_{\rm sc} / Q_{s0}$ can be additionally estimated
using Eq.\,(\ref{eq09}) and the measured bunch parameters.
The particle number and the bunch length could be measured
with a reasonable accuracy.
The transverse beam radius,
which enters the space-charge tune shift as squared
($\varepsilon_y = a_y^2 Q_{0y}/R$, $a_y$ is the vertical rms radius)
and is thus especially important,
could not be determined with a satisfactory precision,
as it was also the case in the previous coasting-beam measurements \cite{paret2010}
at SIS18.
As an example, here we provide
an estimation for the bunch presented in Figs.\,\ref{fg21},\,\ref{fg22}.
The transverse rms emittances were $\epsilon_y=6.2$\,mm\,mrad,
$\epsilon_x=8.4$\,mm\,rad,
number of ions per bunch was 5.1$\times 10^9$.
Using these parameters, the bunch length, and the bare synchrotron tune
(see Table\,1), we obtain from Eqs.\,(\ref{eq09})\,,(\ref{eq10}) $q_{\rm est} \approx$7.
In this work we make no claim on a perfect agreement
of the $q$--values obtained from the transverse spectra
with the  $q$--estimations provided by the bunch parameters and Eq.\,(\ref{eq09}),
mainly due to the uncertainty in the transverse beam size measurements
at SIS18.

\begin{figure}[h!]
\includegraphics*[width=.7\linewidth]{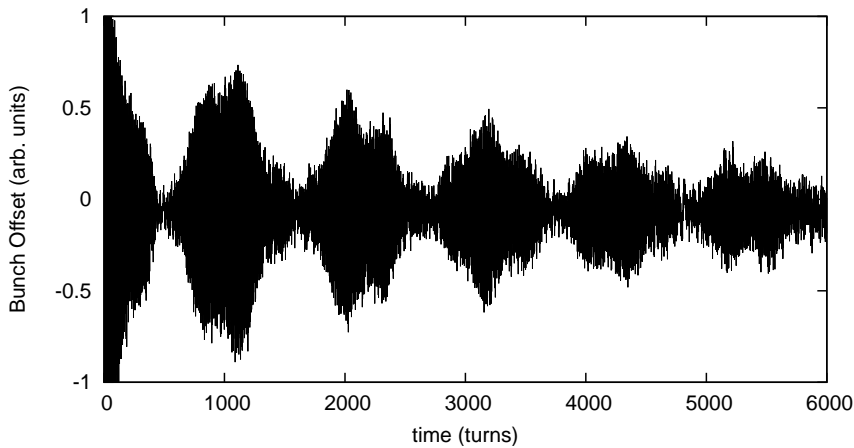}
\caption{\label{fg23}
Time evolution of the bunch offset in the vertical plane
at SIS18 after a transverse kick.
The recoherence periodicity corresponds to the mix of the dominating head-tail modes
$k=0$ and $k=2$ with $\Delta Q_{k=2}=0.91\times10^{-3}$,
giving the periodicity of 1100\,turns.
}
\end{figure}

\begin{figure}[h!]
\includegraphics*[width=.7\linewidth]{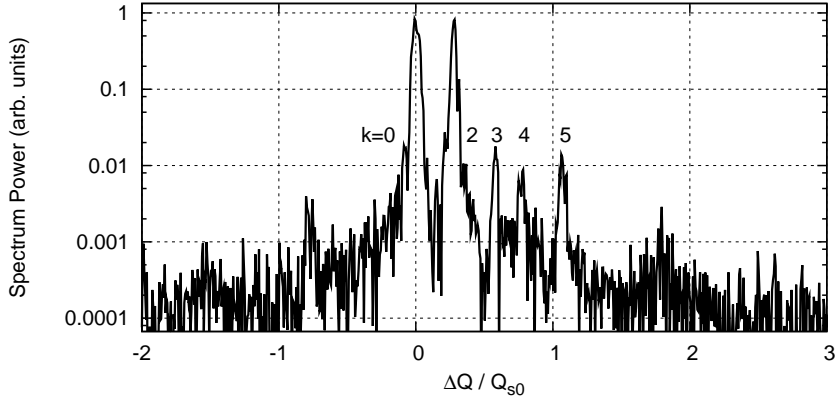}
\caption{\label{fg24}
Transverse coherent spectrum
for the bunch from Fig.\,\ref{fg23}, $Q_{s0}=3.24\times10^{-3}$.
Head-tail modes up to $k=5$ are well seen,
except for the modes $k=1$.
The frequencies of the head-tail modes
provide the space charge parameter $q \approx 9$,
see red squares in Fig.\,\ref{fg30}.
}
\end{figure}

\begin{figure}[h!]
\includegraphics*[width=.7\linewidth]{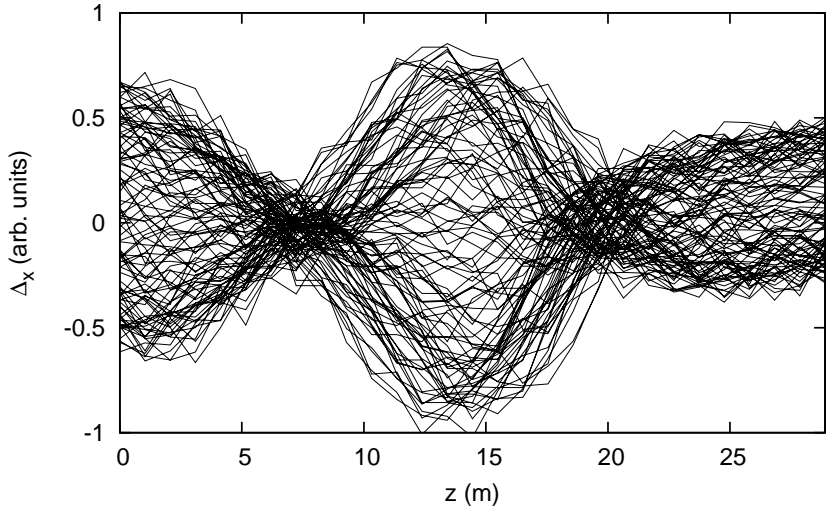}
\caption{\label{fg27}
Traces of the transverse bunch signal for
100 consecutive turns for the bunch from Fig.\,\ref{fg23} and
Fig.\,\ref{fg24}.
This result proves that the mode $k=2$ dominates
during the process of bunch decoherence.
}
\end{figure}

In the next example we show a longer bunch, $\sigma_z=1.15$,
due to a lower rf voltage, see Table\,1.
The transverse bunch oscillations after the kick are shown in Fig.\,\ref{fg23}
and the corresponding spectrum is shown in Fig.\,\ref{fg24}.
In comparison to the previous example (Figs.\,\ref{fg21},\,\ref{fg22}),
the bunch here is longer, but the particle number is higher and
the synchrotron tune is larger, thus the space charge parameter
is similar, $q \approx 9$.
As we can see in Fig.\,\ref{fg24}, the spectrum is dominated
by two modes, the $k=0$ mode at the bare tune,
and another one at $\Delta Q=0.91\times10^{-3}$,
which gives the periodicity of the bunch recoherence,
see Fig.\,\ref{fg23}.
The mode $k=1$ is suppressed as it is the case in the previous example,
and it is to expect that here we have the $k=2$ mode again.
Additionally, this could be proved as follows.
Plotting the bunch vertical traces and subtracting
the total bunch offset, thus reducing the contribution
of the $k=0$ mode, we observe a clear two-knot structure
of the $k=2$ modes, see Fig.\,\ref{fg27}.
The frequencies of the further peaks in Fig.\,\ref{fg24} correspond
rather well to the space-charge model with $q \approx 9$.

\begin{figure}[h!]
\includegraphics*[width=.7\linewidth]{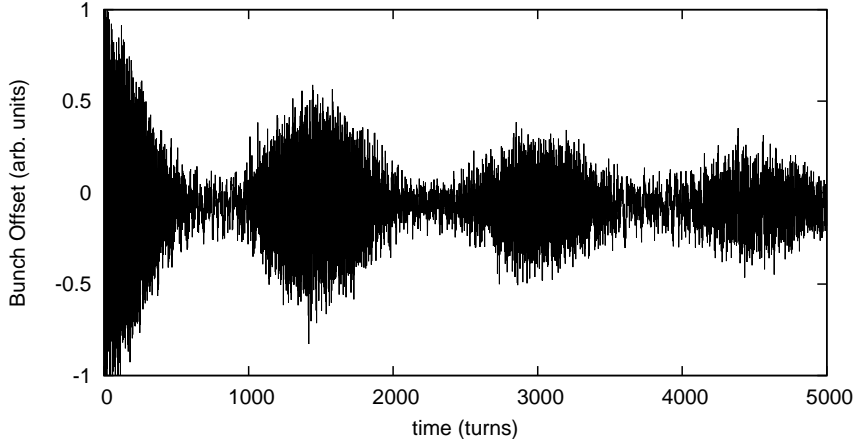}
\caption{\label{fg25}
Time evolution of the bunch offset in the vertical plane
at SIS18 after a transverse kick.
The recoherence periodicity corresponds to the mix of the dominating head-tail modes
$k=0$ and $k=1$ with $\Delta Q_{k=1}=0.68\times10^{-3}$,
giving the periodicity of 1470\,turns.
}
\end{figure}

\begin{figure}[h!]
\includegraphics*[width=.7\linewidth]{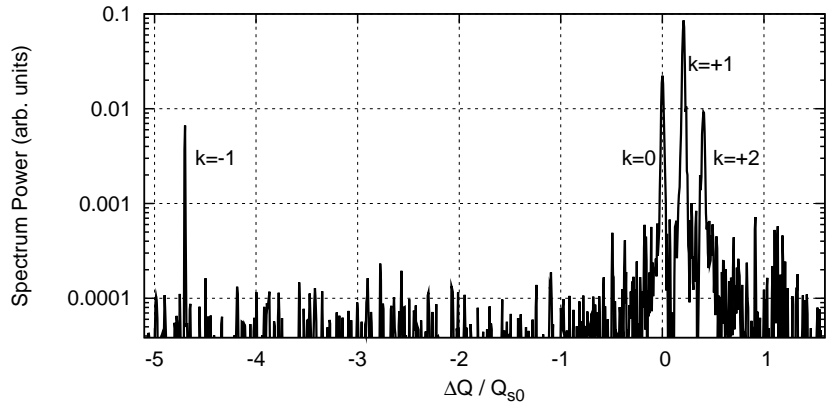}
\caption{\label{fg26}
Transverse coherent spectrum
for the bunch from Fig.\,\ref{fg25}, $Q_{s0}=3.24\times10^{-3}$.
The mode $k=1$ dominates,
the spectrum shows clearly the mode $k=-1$,
with the eigenfrequency corresponding well to the model Eq.\,(\ref{eq05}).
The frequencies of the head-tail modes
provide the space charge parameter $q \approx 4.5$,
see black crosses in Fig.\,\ref{fg30}.
}
\end{figure}

In the next example we demonstrate a bunch decoherence
dominated by a mixture of the $k=0$ mode with the $k=1$ mode;
the bunch oscillations are shown in Fig.\,\ref{fg25},
the spectrum is shown in Fig.\,\ref{fg26}.
The horizontal chromaticity was partly compensated,
by a half of the natural value,
the associated nonlinearities probably contributed to
establishing of the longer bunch
and to a stronger damping of the $k=2$ mode.
The recoherence is thus quite slower, nearly one and a half thousand turns
(see Fig.\,\ref{fg25}),
which is given by the frequency of the $k=1$ mode
in a good agreement with the bunch spectrum, Fig.\,\ref{fg26}.
Another outstanding feature of this spectrum
is the clear presence of the $k=-1$ mode,
with the frequency shifted strongly downwards
in a very good agreement with the space-charge model,
see black crosses in Fig.\,\ref{fg30}.
In part, the presence of the $k=-1$ mode was probably possible due to
rather moderate space charge $q \approx 4.5$ in this case.

\begin{figure}[h!]
\includegraphics*[width=.7\linewidth]{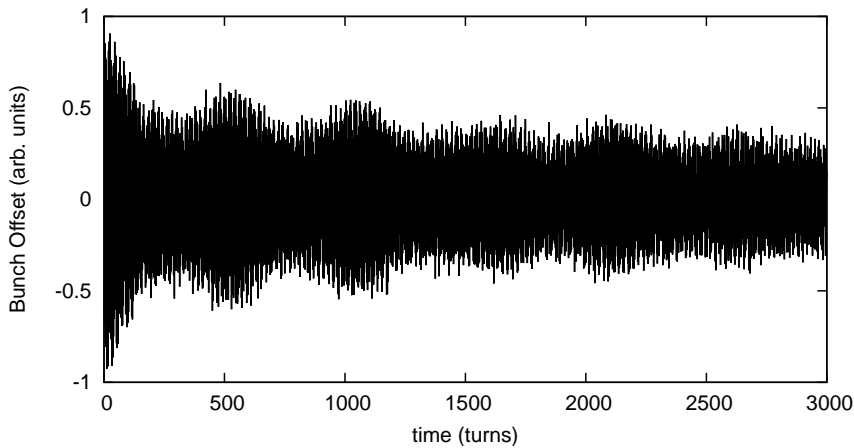}
\caption{\label{fg28}
Time evolution of the bunch offset in the vertical plane
at SIS18 after a transverse kick.
The vertical chromaticity was compensated for this beam to $\xi_y \approx 0$.
The recoherence periodicity corresponds to the mix of the dominating head-tail modes
$k=0$ and $k=2$ with $\Delta Q_{k=2}=1.86\times10^{-3}$,
giving the periodicity of 540\,turns.
}
\end{figure}

\begin{figure}[h!]
\includegraphics*[width=.7\linewidth]{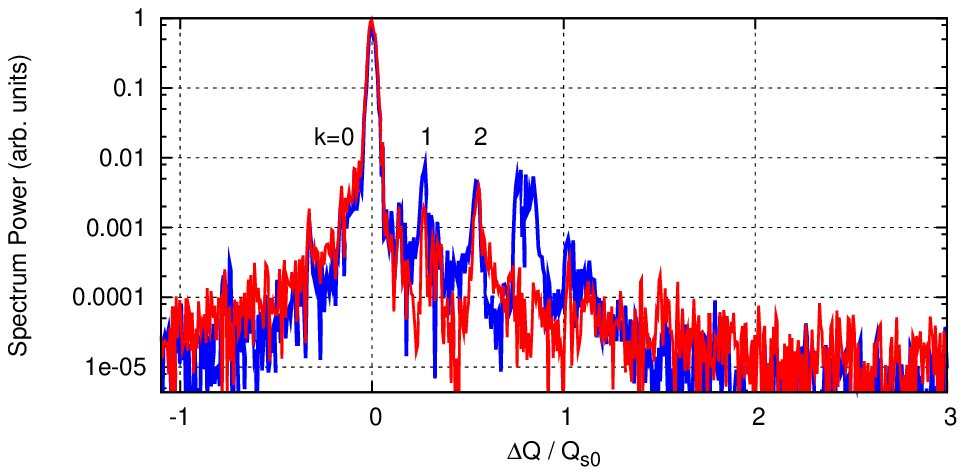}
\caption{\label{fg29}
Transverse coherent spectrum
for the bunch from Fig.\,\ref{fg28}, $Q_{s0}=3.24\times10^{-3}$.
The red spectrum is obtained from a frequency analysis
of the complete bunch offset,
while the blue spectrum is a result of a frequency analysis
for the bunch head.
The mode $k=0$ highly dominates,
in the complete-bunch spectrum the mode $k=2$ is stronger,
and the bunch head spectrum reveals the uneven modes
$k=1$, $k=3$.
The frequencies of the head-tail modes
provide the space charge parameter $q \approx 4$,
see cyan triangles in Fig.\,\ref{fg30}.
}
\end{figure}

The transverse decoherence observed in the case presented
in Figs.\,\ref{fg28},\,\ref{fg29}, is very different
from the third example, Fig.\,\ref{fg25},
although the space-charge parameter is similar, $q \approx 4$,
as well as the bunch length, see Table\,1.
We see that the recoherence periodicity is quite faster
which is due to the dominance of the $k=2$ mode
as it is confirmed in the bunch spectrum, see the red line
in Fig.\,\ref{fg29}.
More remarkable, the bunch decoherence in Fig.\,\ref{fg28}
shows a much weaker amplitude drop between the recoherence peaks.
The reason is the compensated vertical chromaticity to nearly zero,
according to the set parameters.
This is predicted by the linear theory Eq.\,(\ref{eq04}),
also shown in Fig.\,\ref{fg08}.
According to the interpretation of the mode mixture,
at a small chromaticity the part of the $k=0$ mode is very large,
see Fig.\,\ref{fg09}.
The spectrum from the measurements in Fig.\,\ref{fg29} confirms this.
The relatively small part of the $k=2$ mode provides
the periodicity of a weak recoherence.
For the determination of the space-charge parameter,
the eigenfrequency of the $k=1$ mode is needed
which could be obtained by a frequency analysis of the
bunch head oscillations, see the blue line in Fig.\,\ref{fg29},
and the resulting $q$--values in Fig.\,\ref{fg30} (cyan triangles).

\section{CONCLUSIONS}

The transverse decoherence and coherent eigenspectra
in long bunches with space charge have been studied using measurements at the SIS18 heavy-ion synchrotron and
particle tracking simulations.

A model Eq.\,(\ref{eq05}) for the combined effect of space charge and nonlinear synchrotron oscillations
has been formulated, with the fitting parameter $q_*$ obtained from
the particle tracking simulations for the low-order head-tail modes.
The space-charge parameter $q=\Delta Q_{\rm sc}/Q_{s0}$
of the bunch can be determined for every head-tail mode
from the corresponding frequency shift $\Delta Q_k$, see Eq.\,(\ref{eq06}),
according to the given bunch length.

The transverse decoherence in bunches with space charge has been observed experimentally 
and quantitively explained, using simulations and analytic models. 
An initial rigid bunch offset can be decomposed into
head-tail modes. The chromaticity determines
the contribution of the different head-tail modes.
Using the airbag \cite{blask98} eigenmodes as an approximation for
the bunch head-tail modes, the relative amplitudes
can be found analytically, see Fig.\,\ref{fg07}.

Different head-tail modes experience also different Landau damping rates.
After a transition period the bunch oscillation
is a combination of the remaining modes.
For example, it can be a mix of the $k=0$ mode and $k=1$ mode.
The periodicity of the bunch recoherence corresponds then
to the frequency difference between these two modes.
Our simulation examples demonstrate this explanation of the bunch decoherence
for different space-charge parameters and for different chromaticities,
see Figs.\,\ref{fg06},\,\ref{fg07}.
In the simulations the dominating Landau damping mechanism
is due to the variation of the space charge tune shift along the bunch
\cite{burov2009, balb2009, kornilov_prstab10}.

Experimental observations of the transverse bunch decoherence with space charge in the SIS18 
heavy-ion synchrotron at GSI are presented.
The space charge parameter $q$ has been determined from the
bunch spectra for different head-tail modes,
summarized in Fig.\,\ref{fg30}.
With increasing bunch length we observe that nonlinear synchrotron oscillations  
modify the head-tail mode frequencies.
The bunch decoherence always corresponded to the mix of
the dominating modes, in our case the $k=0$ and $k=1$ modes
or the $k=0$ and $k=2$ modes.
Compared to the simulation it is more difficult to predict which modes
would be faster suppressed due to additional damping mechanisms
in a real machine.
In the experiment the oscillations are further damped after the transition period,
possibly due to the nonlinear magnet field errors.
The periodicity of the recoherence was
exactly confirmed by the mode frequencies from the spectra.
A direct comparison of the first two examples
(Figs.\,\ref{fg21},\,\ref{fg22} vs.\ Figs.\,\ref{fg23},\,\ref{fg24})
demonstrates the role of the bunch length.
A comparison of the the fourth example (Figs.\,\ref{fg28},\,\ref{fg29})
with the others demonstrates the role of the chromaticity:
at a nearly zero chromaticity the mode $k=0$
dominates the bunch decoherence alone.
The third example (Figs.\,\ref{fg25},\,\ref{fg26})
shows a decoherence with pronounced flat 
between the recoherence peaks, corresponding to the
mix of the $k=0$ and $k=1$ modes.

The results of this work apply to the evolution of a possible transverse
injection offset during bunch-to-bucket
transfer from one ring to another.
Transverse coherent spectra can be used not only
to measure the betatron tune, the head-tail mode frequencies
can be used to extract useful information about the long intense bunches.
The understanding of the decoherence with space charge
and rf nonlinearity
has direct consequences for the chromaticity
measurements in intense bunches, which should be
analysed in detail in a future work.

\end{document}